\documentclass[10pt,twocolumn,american]{revtex4-1}
\usepackage[T1]{fontenc}
\usepackage{babel}
\usepackage{amstext}
\usepackage{graphicx}
\usepackage[unicode=true,
 bookmarks=true,bookmarksnumbered=false,bookmarksopen=false,
 breaklinks=true,pdfborder={0 0 0},backref=false,colorlinks=false]
 {hyperref}
\begin{document}

\title{img2net: Automated network-based analysis of imaged phenotypes}

\author{David Breuer$^{1,*}$ and Zoran Nikoloski$^{1}$}

\affiliation{$^{1}$Mathematical Modeling and Systems Biology, Max Planck Institute
of Molecular Plant Physiology, Am Muehlenberg 1, 14476 Potsdam-Golm,
Germany}

\affiliation{$^{*}$\href{mailto:breuer@mpimp-golm.mpg.de}{breuer@mpimp-golm.mpg.de}}
\begin{abstract}
Summary: Automated analysis of imaged phenotypes enables fast and
reproducible quantification of biologically relevant features. Despite
recent developments, recordings of complex, networked structures,
such as: leaf venation patterns, cytoskeletal structures, or traffic
networks, remain challenging to analyze. Here we illustrate the applicability
of img2net to automatedly analyze such structures by reconstructing
the underlying network, computing relevant network properties, and
statistically comparing networks of different types or under different
conditions. The software can be readily used for analyzing image data
of arbitrary 2D and 3D network-like structures. 

~

Availability and Implementation: img2net is open-source software under
the \href{http://www.gnu.org/copyleft/gpl.html}{GPL} and can be downloaded
from \href{http://mathbiol.mpimp-golm.mpg.de/img2net/}{http://mathbiol.mpimp-golm.mpg.de/img2net/},
where supplementary information and data sets for testing are provided.

~

\end{abstract}
\maketitle

\section{Introduction}

Biological and man-made systems, ranging from biochemical reactions
to neural and social interactions, can often be represented as networks,
with nodes and edges representing the components and their interactions,
respectively. Network representations facilitate not only intuitive
visualization, but also quantitative studies of the systems' structure
and dynamics. 

Spatial networks constitute an import subclass which includes networks,
such as: leaf venation \cite{Dodds2010}, cellular cytoskeleton \cite{Volkmann1999},
and city streets \cite{Barthelemy2011}. While some of these networks,
like city infrastructure, have been well-characterized, others, like
biological spatial networks, remain poorly understood.

Snapshots of spatial networks can be obtained by imaging technologies.
The computational challenge is that of extracting the the underlying
networks from the gathered images in a fast and reliable fashion,
and of examining the reconstructions to reveal the underlying organizational
principles. 

The existing image-based methods for reconstruction of biological
networks are typically designed for specific types of networks: Tree-like
networks for plant root architectures \cite{pound2013rootnav}; fungal
or leaf venation networks \cite{Obara2012a}; or neuronal topology
of the human connectome \cite{meijering2010neuron,Longair01092011}.
While most of these approaches require user input, rendering them
unfeasible for high-throughput studies, fully automated algorithms
are usually tailored to specific image sources and challenged by low
signal-to-noise ratios, which may strongly affect the resulting networks.

Here, we present a robust approach to reconstruct 2D and 3D (non-)biological
spatial networks from gray-scale image data. By constructing weighted
networks and computing their seminal network properties, img2net allows
an extensive quantification and statistical comparison of network
topologies.

\section{Methods and Functionality}

We extract the networks from image data in two steps: Starting from
a gray-scale image, where high intensities reflect strong links in
the network, we place a (arbitrary) grid which covers the region of
interest. For each grid-edge, we determine a weight by convolving
a Gaussian kernel with the original image to capture the edge's capacities
to carry certain traffic, the speed of transportation, or combinations
thereof (cf.~Results). 

To facilitate comparative analyses, different seminal properties of
the resulting weighted network are computed, including: degree distribution,
path lengths, or random-walk-related properties. To evaluate the biological
importance of the calculated properties, we developed several null
models which randomize the network while preserving the distribution
of edge weights. They provide the basis for revealing the principles
underlying the network organization \cite{breuer2014grid}. 

img2net is written in Python and provides a graphical user interface
(GUI; Fig.~\ref{fig:fig}A) for the selection of the image input
data and the parameters of the network reconstruction procedure. It
operates on folders of~.png,~.jpg, or~.tiff files that display
the networked structure of interest, with a gray-scale representation
of edge strengths. The directory tree divides the images into different
treatments and experiments, e.g.:
\[
\textrm{root/treatment\_003/experiment\_005/image\_z001\_t001.tiff}
\]
The usage of rectangular, triangular, or hexagonal grids is supported;
grid spacings, periodic boundary conditions, the widths of the convolution
kernel and the number of layers for 3D image data can be specified
by the user. Different null models are available and the number of
null model realizations can be set. Finally, img2net can be run on
multiple cores.

An output folder is generated in the root directory of the chosen
image directory tree: A human-readable document is created which contains
the values of the network properties for the analyzed networks and
their null models to allow further analyses and visualizations of
the results. img2net generates standard plots as~.svg files: For
each experiment, a reconstructed network overlaying the original image
data is saved. For all analyzed network properties, plots of the resulting
time series are generated to facilitate network comparisons.

\section{Results}

\begin{figure}
\begin{centering}
\includegraphics[width=1\columnwidth]{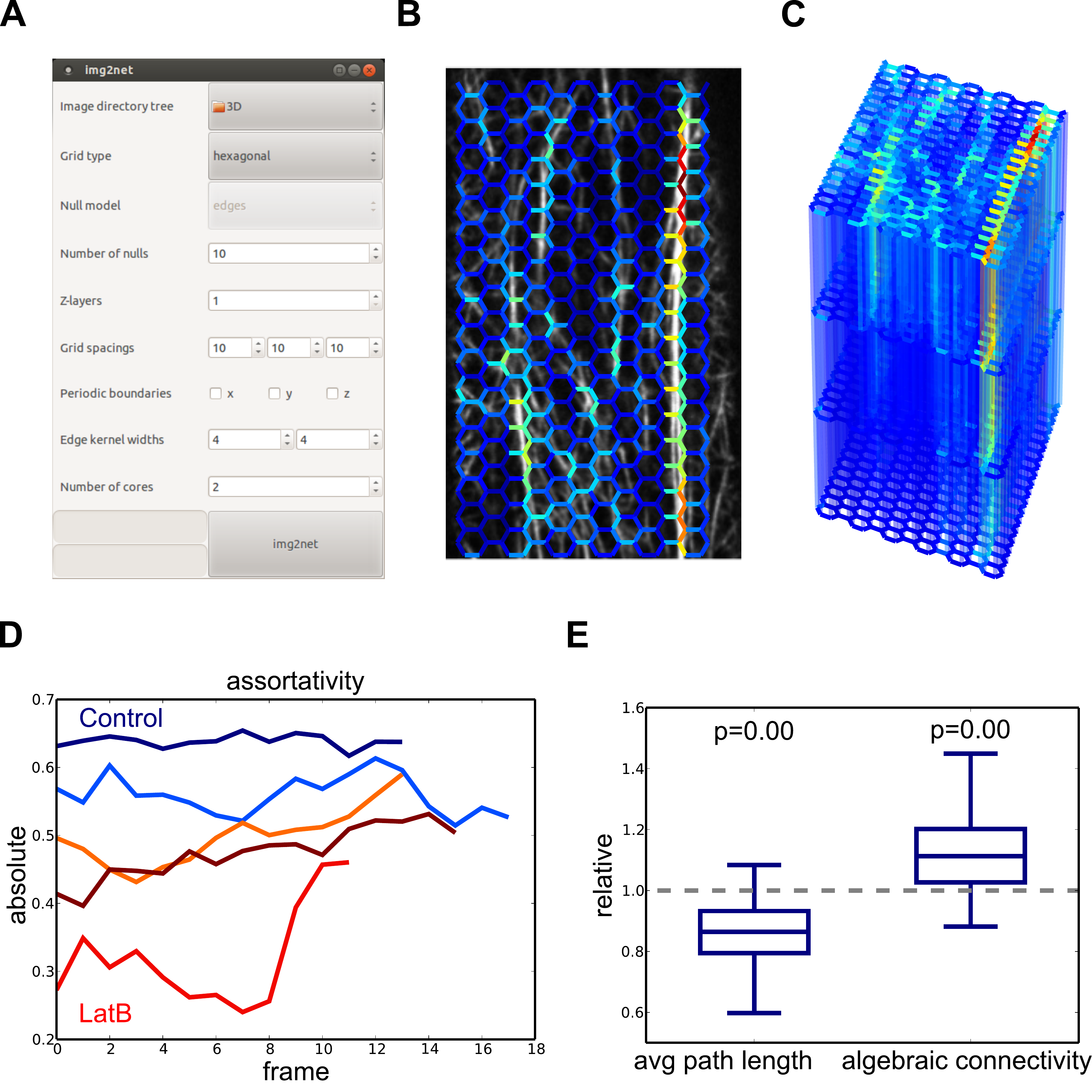}
\par\end{centering}

\protect\caption{\textbf{\label{fig:fig}Img2net: graphical user interface and output
visualizations.} \textbf{(A)} GUI to set parameters for the network
reconstruction. \textbf{(B)} Actin network reconstructed with a hexagonal
grid. \textbf{(C)} 3D reconstruction for image data with four z-slices.
\textbf{(D)} Time series of the assortativity for multiple untreated
(blue) and treated (red) plants. \textbf{(E)} Ratios of average path
lengths and algebraic connectivities of observed and null model networks.}

\end{figure}

For testing, img2net was used to compare confocal recordings of the
actin cytoskeleton (Fig.~\ref{fig:fig}B and C) of untreated plant
cells and cells treated with an actin disrupting drug \cite{breuer2014grid}.
We find that the actin networks of treated cells display a consistently
lower assortativity (Fig.~\ref{fig:fig}D), indicating a drug-related
filament fragmentation. Further, for untreated cells, the observed
networks display significantly smaller average path lengths (\textquotedblleft accessibility\textquotedblright )
and significantly higher algebraic connectivities (\textquotedblleft robustness\textquotedblright )
than expected by chance ($p$-values estimated from the null model
via one-sample $t$-tests; Fig.~\ref{fig:fig}E), suggesting a biological
basis for the maintenance of short and robust transportation routes. 

We also used img2net to analyze the network structure of the German
autobahn which was obtained from OpenStreetMap as an image by using
a gray-scale coding of the speed limits \cite{breuer2014grid}. We
find that, similar to the actin cytoskeleton, the autobahn shows significantly
smaller path lengths and a higher algebraic connectivity than expected
by chance.

\section{Conclusion}

Since networks provide an intuitively accessible as well as mathematically
sound framework for the representation of complex systems, the reconstruction
of spatial networks from image data is useful in biological and technical
research. img2net implements an automated method for fast and robust
reconstruction of arbitrary 2D and 3D (non-)biological spatial networks. 

As our approach relies on fixed grid topologies, img2net offers different
types of grids to verify that the findings are grid-independent. For
small grid spacings, the grid typically approaches the \textquotedblleft true\textquotedblright{}
structure of the underlying network. The main benefit of this approach
is its robustness against noise and flaws in the input images. For
example, small ruptures in the underlying network do not disrupt the
corresponding edges but only weaken them. Hence, our approach does
not require sophisticated error-correcting image processing steps.
Furthermore, img2net is directly applicable to a wide range of image
data from different sources.

The analyses of the reconstructed networks implemented in img2net
allow a quantification of structural properties, comparisons of networks,
e.g., under different conditions, and an assessment whether or not
the network properties reflect underlying organizational principles.

\section{Acknowledgement}

We thank Alexander Ivakov and Staffan Persson for help with the biological
experiments and discussions on cytoskeletal networks. 

Funding: DB and ZN were supported by the Max Planck Society.

Conflict of Interest: none declared.

\bibliographystyle{apalike}

\begin{thebibliography}{}

\bibitem[Barth{\'e}lemy, 2011]{Barthelemy2011}
Barth{\'e}lemy, M. (2011).
\newblock Spatial networks.
\newblock {\em Phys Rep}, 499(1):1--101.

\bibitem[Breuer et~al., 2014]{breuer2014grid}
Breuer, D., Ivakov, A., Sampathkumar, A., Hollandt, F., Persson, S., and
  Nikoloski, Z. (2014).
\newblock Quantitative analyses of the plant cytoskeleton reveal underlying
  organizational principles.
\newblock {\em J R Soc Interface}, 11(97):20140362.

\bibitem[Dodds et~al., 2010]{Dodds2010}
Dodds, P.~S., Corson, F., Katifori, E., Sz{\"o}ll{\H{o}}si, G.~J., and
  Magnasco, M.~O. (2010).
\newblock Focus: why leaves aren't trees.
\newblock {\em Phys Rev Focus}, 25:4.

\bibitem[Longair et~al., 2011]{Longair01092011}
Longair, M.~H., Baker, D.~A., and Armstrong, J.~D. (2011).
\newblock Simple neurite tracer: open source software for reconstruction,
  visualization and analysis of neuronal processes.
\newblock {\em Method Biochem Anal}, 27(17):2453--2454.

\bibitem[Meijering, 2010]{meijering2010neuron}
Meijering, E. (2010).
\newblock Neuron tracing in perspective.
\newblock {\em Cytom Part A}, 77(7):693--704.

\bibitem[Obara et~al., 2012]{Obara2012a}
Obara, B., Grau, V., and Fricker, M.~D. (2012).
\newblock A bioimage informatics approach to automatically extract complex
  fungal networks.
\newblock {\em Bioinformatics}, 28(18):2374--2381.

\bibitem[Pound et~al., 2013]{pound2013rootnav}
Pound, M.~P., French, A.~P., Atkinson, J., Wells, D.~M., Bennett, M.~J., and
  Pridmore, T.~P. (2013).
\newblock Root{N}av: navigating images of complex root architectures.
\newblock {\em Plant Physiol}, 162(4):1802--1814.

\bibitem[Volkmann et~al., 1999]{Volkmann1999}
Volkmann, D., Baluska, F.~E., et~al. (1999).
\newblock Actin cytoskeleton in plants: from transport networks to signaling
  networks.
\newblock {\em Microsc Res Tech}, 47(2):135--154.

\end{thebibliography}

\end{document}